\documentclass[a4paper,10pt,twocolumn]{article}
\usepackage{inputenc}
\usepackage{tgtermes}
\usepackage{titlesec}
\usepackage{float}
\titleformat{\section}
{\normalfont\Large}{\thesection}{1em}{}
\titleformat{\subsection}
{\normalfont\large}{\thesubsection}{1em}{}

\usepackage{soul}

\usepackage{amsmath,amssymb}
\usepackage{mathptmx}
\usepackage{upgreek}
\usepackage{nicefrac}

\usepackage{graphicx}

\usepackage{xspace}

\usepackage[super, comma, sort&compress]{natbib}
\newcommand{\ket}[1]{\ensuremath{|#1\mkern-1mu\rangle}}

\newcommand{\um}{\upmu \text{m}}

\newcommand{\bra}[1]{\ensuremath{\langle #1 \mkern-1mu|}}

\newcommand{\Mthree}{\ensuremath{\mathcal{M}_{I\mkern -2mu I\mkern -2mu I}^G}\xspace}
\newcommand{\Mtwo}{\ensuremath{\mathcal{M}_{I\mkern -1mu I}^G}\xspace}

\newcommand{\Nmax}{N_\mathrm{max}}
\newcommand{\Nmin}{N_\mathrm{min}}

\newcommand{\Mtol}{\ensuremath{\mathcal{M}_{I\mkern -1mu I}^L}\xspace}
\newcommand{\Mtos}{\ensuremath{\mathcal{M}_{I\mkern -1mu I}^S}\xspace}

\newcommand{\Mtolvar}{\ensuremath{\mathcal{M}_{I\mkern -1mu I'}^{L}}\xspace}
\newcommand{\Mtosvar}{\ensuremath{\mathcal{M}_{I\mkern -1mu I'}^{S}}\xspace}

\usepackage[usenames, dvipsnames]{color}
\newcommand{\josh}[1]{{\color{Blue}#1}}
\newcommand{\jez}[1]{{\color{Red}#1}}

\renewcommand{\josh}[1]{{#1}}
\renewcommand{\jez}[1]{{#1}}

\usepackage{authblk}

\title{Programmable four-photon graph states on a silicon chip}
\date{}

\author[]{Jeremy C. Adcock}
\author[]{Caterina Vigliar}
\author[]{Raffaele Santagati}
\author[]{\\Joshua W. Silverstone\thanks{josh.silverstone@bristol.ac.uk}}
\author[]{Mark G. Thompson}
\affil[]{\small{Quantum Engineering Technology (QET) Labs, H. H. Wills Physics Laboratory \& School of Computer, Electronic Engineering \& Engineering Mathematics, University of Bristol, Merchant Venturers Building, Woodland Road, Bristol BS8 1UB, UK}}

\begin{document}

\maketitle

{\bfseries
Future quantum computers require a scalable architecture on a scalable technology---one that supports millions of high-performance components.
Measurement-based protocols, based on graph states, represent the state of the art in architectures for optical quantum computing \cite{raussendorf2001one, rudolph2017optimistic, gimeno2015three}. 
Silicon photonics offers enormous scale\cite{sun2013large, chung2018monolithically} and proven quantum optical functionality\cite{silverstone2016silicon}.
Here we report the first demonstration of photonic graph states on a mass-manufactured chip using four on-chip generated photons.
We generate both star- and line-type graph states, implementing a basic measurement-based protocol, and measure heralded interference of the chip's four photons.
We develop a model of the device and bound the dominant sources of error using Bayesian inference.
The two-photon barrier, which has constrained chip-scale quantum optics, is now broken; future increases in on-chip photon number now depend solely on reducing loss, and increasing rates.
This experiment, combining silicon technology with a graph-based architecture, illuminates one path to a large-scale quantum future.
}

\begin{figure*}[t!]
\centering
\includegraphics[width=1\linewidth]{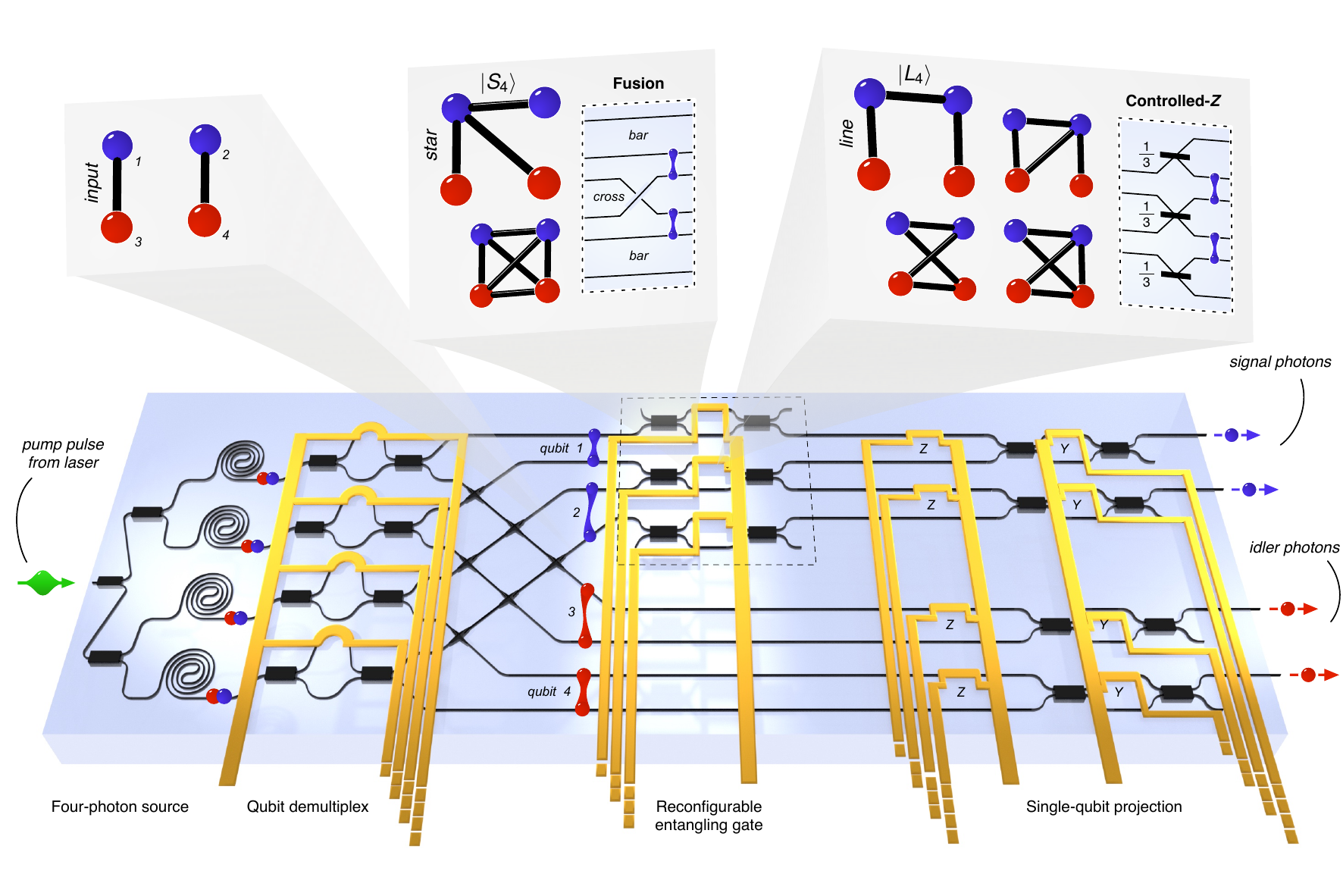}
\caption{Experiment overview. A schematic of the silicon-on-insulator chip-scale device is shown, comprising: four telecommunications-band photon pair sources, producing four photons in superposition; a qubit demultiplexer, which configures that superposition into a product of two Bell pairs; a reconfigurable postselected entangling gate (R-PEG); and four single-qubit projection and analysis stages, formed of four Mach-Zehnder interferometers implementing qubit $Y$ rotations, preceded by four $Z$ rotations. An optical micrograph of the device can be found in the Supplementary Information. Corresponding graph states are indicated above, starting with the two input Bell pairs, and ending with either `star' or `line' graph states, for fusion or controlled-$Z$ R-PEG configurations, respectively.}
\label{fig-overview}
\end{figure*}

Graph states are key entangled resources  for quantum information processing. They are quantum states which can be drawn as a graph, with a qubit on each vertex and \josh{local entanglement} on each edge\cite{hein2004multiparty}. In measurement-based quantum computing, \josh{where single-qubit measurements on a graph state drive the computation forward,} particular graphs enable particular computational tasks \cite{hein2006entanglement}.
Topological quantum error correction, relying centrally on graph states, will provide essential noise tolerance to future experimental realisations\cite{RaussendorfErrorCorr}. Graph states also play a central role as platforms for the simulation of complex processes and dynamics\cite{SimulationGraphsGeorgescu}, and for quantum secret sharing protocols\cite{markham2008graph}. 
As such, graph states have featured strongly in experiment, in both optics\cite{walther2005experimental2, bell2014experimental2, ciampini2016path} and other platforms\cite{wang201816}. The reconfigurable generation of \textit{arbitrary} graphs, never before achieved in optics, will accelerate development of many graph-based applications. 

Integrated optics promises new levels of scale for optical quantum devices. It offers robustly mode-matched, miniature components, lithographically defined in a planar process. State-of-the-art chip-scale devices exhibit \josh{loss and error} performance approaching that of bulk and fibre systems. Quantum optical functionality has been demonstrated in all major technology platforms: lithium niobate\cite{alibart2016quantum}, silica\cite{politi2008silica, crespi2016suppression, spring2017chip} (both lithographic and laser-written), silicon nitride\cite{taballione2018programmable}, gallium arsenide\cite{dietrich2016gaas}, indium phosphide\cite{sibson2017chip}, and silicon\cite{silverstone2016silicon, silverstone2014on}.

Silicon devices have rapidly grown in complexity in recent years, with quantum demonstrators now exceeding 500 on-chip components\cite{wang2018multidimensional}, and classical silicon photonic devices having thousands\cite{sun2013large, chung2018monolithically}. Integration with CMOS electronics could push this scale further still, \josh{by miniaturising control and interconnect functionality}\cite{chung2018monolithically}. \josh{A device's computational power is related to the quantum configuration---or Hilbert---space accessible to it. In optics, this space has $m^n$ dimensions, for $n$ photons scattered across $m$ modes. So far, the scaling up of silicon \textit{quantum} photonics has involved scattering only one or two photons ($n=1$ or $2$) over more and more waveguides (increasing $m$) as a route to larger Hilbert spaces\cite{harris2017pnp, wang2018multidimensional}.} Only recently has on-chip heralded interference between on-chip-generated photons been demonstrated\cite{faruque2018hom}, though visibility is limited and no quantum information has yet been encoded. Extending chip-scale quantum optics into the multi-pair regime, increasing $n$, is a crucial step.

We present a silicon quantum-optical device (Fig.~\ref{fig-overview}) which can generate four photons and use them to prepare both \josh{classes\cite{adcock2018hard} of four-qubit graph state entanglement---classes} closed under local unitary transformations. We refer to these classes by their best-known members: `star' $\ket{S_4}$, and `line' $\ket{L_4}$. Our device operates in four stages. (\textbf{1})~Four photons in two pairs are generated in superposition over four sources. (\textbf{2})~These are demultiplexed by wavelength and rearranged to group signal and idler photons. The resulting dual-rail, path-encoded \josh{qubit} state is a product of Bell pairs, $\ket{\Phi^+}_{1,3}\otimes \ket{\Phi^+}_{2,4}$ (with qubit indices in subscript). (\textbf{3})~The \josh{signal-photon qubits} are operated upon by a reconfigurable postselected entangling gate (R-PEG). This can be programmed to perform either a fusion or controlled-$Z$ operation, to generate star- or line-type entanglement\cite{adcock2018hard}, respectively, with postselected probability $\nicefrac{1}{2}$ or $\nicefrac{1}{9}$. (\textbf{4})~We then perform arbitrary single-qubit projective measurements, using Mach-Zehnder interferometers (MZI), on the four-qubit states. \jez{A full description of the state evolution is in the Supplementary Information.}

The $\chi^{(3)}$ process, spontaneous four-wave mixing, converts bright telecommunications-band pump pulses into quantum-correlated signal and idler photons in the spiralled silicon waveguides of our source stage\cite{sharping2006generation}. Thermo-optic phase modulators provide electronic reconfigurability throughout the device. Focussing vertical grating couplers connect on-chip waveguides to optical fibre. Finally, signal and idler photons are tightly filtered in fibre (pump:photon filtering ratio 2:1), and registered by superconducting nanowire single-photon detectors. \josh{See Methods for more details.}

\begin{figure*}[t!]
\centering
\includegraphics[width=1\linewidth]{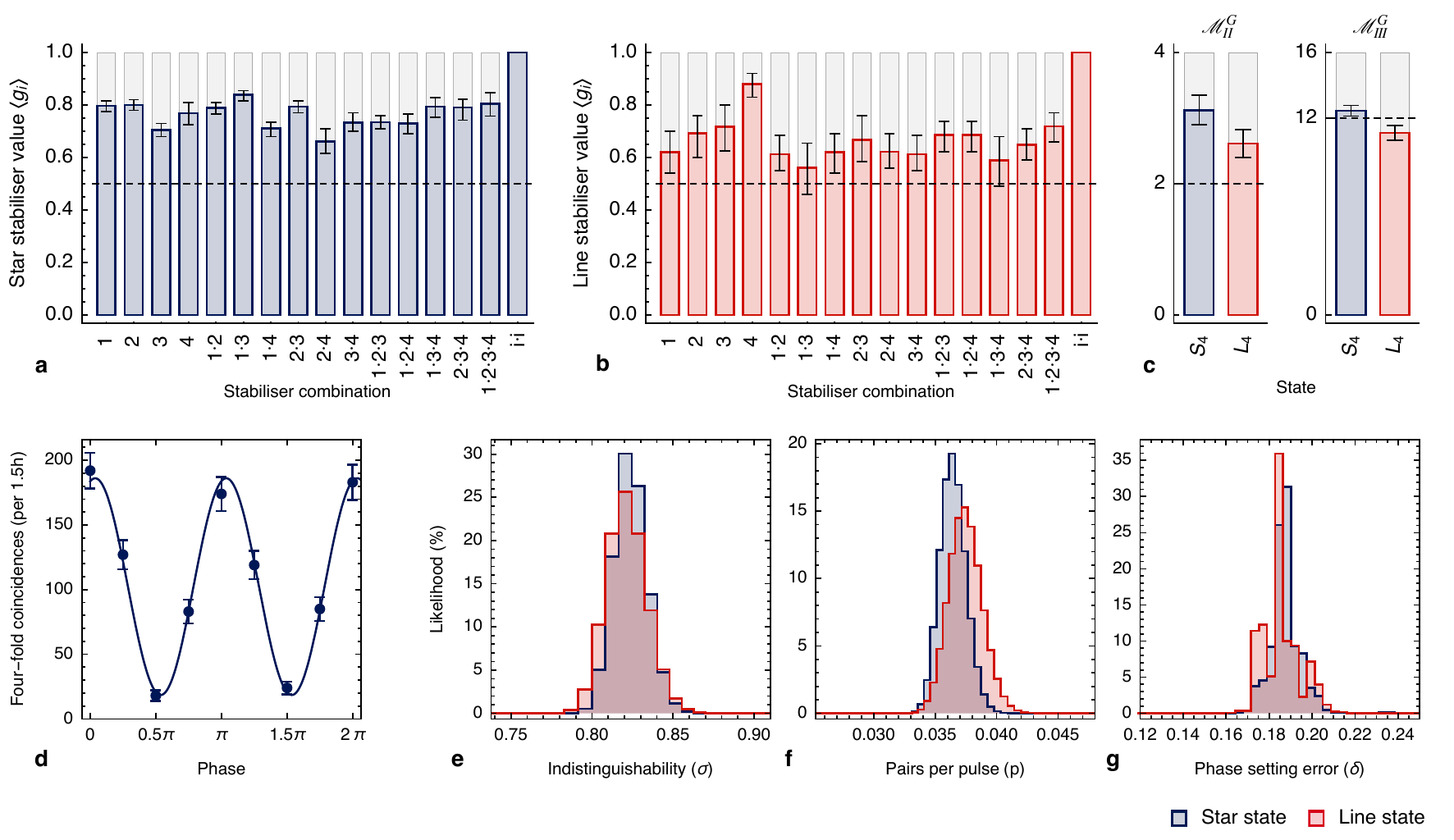}
\caption{Summary of experimental data. \textbf{a,b.} Stabiliser observables of the star and line graph states, $\bra{S_4}g_i\ket{S_4}$ and $\bra{L_4}g_i\ket{L_4}$, used to estimate state fidelity. Dashed lines indicate the $F>\nicefrac{1}{2}$ threshold for witnessing genuine multipartite entanglement, where $F=\mathrm{mean}\{\langle{g_i}\rangle\}$. \textbf{c.} Mermin parameters \Mtwo and \Mthree for the $\ket{S_4}$ and $\ket{L_4}$ states, estimated from stabilisers measurements. Local hidden variable bounds are indicated with a dashed line. Values are reported in Table~\ref{tab-data}. \textbf{d.} On-chip Hong-Ou-Mandel interference, with \josh{HOM fringe visibility of $V = 0.82 \pm 0.02$}. Probability distributions for the (\textbf{e}) indistinguishability, (\textbf{f}) source brightness, and (\textbf{g}) phase error, derived via a Bayesian parameter estimation method.}
\label{fig-data}
\end{figure*}

High-quality two-photon interference between on-chip-generated photons is central to the device's operation. We measure heralded Hong-Ou-Mandel (HOM) interference, between the signal photons of sources $2$ and $3$, on an \josh{R-PEG} MZI, heralded by the idler photons. We find a visibility of $V = (\Nmax-\Nmin)/(\Nmax+\Nmin) = 0.82\pm 0.02$, where $\Nmax$ and $\Nmin$ are the maximum and minimum values of the fitted sinusoid. Shown in Fig.~\ref{fig-data}d, this is the highest such visibility measured on a chip to date\cite{faruque2018hom}. The conventional HOM-dip-equivalent visibility, equivalent to the heralded purity, is $V_{\mathrm{HOM}} = (\Nmax-2 \Nmin)/\Nmax =  0.80\pm0.02$ (see Supplementary Information). The photon-pair generation probability here is $p = 0.06$. We corroborate $V$ by measuring the unheralded \josh{second-order correlation function} $g^{(2)}(0)$ for the eight modes of our four on-chip sources, implying\cite{christ2011probing} heralded purities between $0.82$ and $0.92$ \jez{(the Supplementary Information contains a full listing)}. \josh{New parametric source designs will further improve brightness and purity on-chip}\cite{spring2017chip, christensen2017pump, vernon2017amzimrr}.

We verify the generation of the four-photon star and line graph states ($\ket{S_4}$ and $\ket{L_4}$) by measuring their 16 stabilisers\cite{guhne2009entanglement}, $g_{\{i\}}$, where $\{i\}$ is the set of generators whose product composes each stabiliser (e.g. $g_{12} = g_1 g_2$).  The four stabiliser generators of the state $\ket{S_4}$ are:
\begin{equation*}
    \begin{split}
    g_1 = XIIZ,& \  g_2 = IXIZ,\\  g_3 = IIXZ,&  \ g_4 = ZZZX,
    \end{split}
\end{equation*}
where $X$, $Y$, and $Z$ are Pauli matrices and $I$ is the identity matrix; tensor products are implied. For $\ket{L_4}$, the stabiliser generators are:
\begin{equation*}
    \begin{split}
    g_1 &= XZZI,  \ g_2 = ZXIZ,\\ g_3 &= ZIXI,  \ g_4 = IZIX.
    \end{split}
\end{equation*}
The measured stabilisers are plotted in Figs~\ref{fig-data}a and \ref{fig-data}b, for $\ket{S_4}$ and $\ket{L_4}$ respectively. From these, we compute fidelities, shown in Table~\ref{tab-data}, and find that both states robustly satisfy the $F>\nicefrac{1}{2}$ threshold for witnessing genuine multipartite entanglement\cite{guhne2009entanglement}.  These fidelities compare favourably with previous bulk-optics measurements on these states\cite{zhao2003experimental, zhang2016experimental}. In these and subsequent four-photon measurements, we reduce the photon-pair generation probability to $p = 0.03$ \josh{to suppress multiphoton contamination.}

We perform a basic measurement-based protocol\cite{raussendorf2001one} by projecting various qubits of $\ket{S_4}$ onto $\ket{0}$, and measuring the remaining two- and three-qubit graph states. We denote these states $ \ket{S_4}_{J} = \big(\bigotimes_{j\notin J} \bra{0}_j\big)\ket{S_4} $, where $J$ is the set of remaining \josh{(un-projected)} qubits. The three-qubit state $\ket{S_4}_{1,2,4}$ and the two-qubit states $\ket{S_4}_{1,4}$ and $\ket{S_4}_{3,4}$ can be produced by projecting qubits \{3\}, \{2,3\}, and \{1,2\} onto $\ket{0}$, respectively. \josh{Measured fidelity data for these states are listed in Table~\ref{tab-data}.} Notice that the two photons encoding $\ket{S_{4}}_{1,4}$ are orthogonal in colour, and have never interacted.

Mermin tests let us verify the nonlocality of multipartite states\cite{walther2005experimental, CaterinaMultipartiteTest}. \josh{We construct tests\cite{guhne2009entanglement} comprising two and three measurement settings per qubit, \Mtwo and \Mthree, based on the stabiliser observables of each graph state $G$. Results are listed in Table~\ref{tab-data} and plotted in Fig.~\ref{fig-data}c. \jez{\Mtwo allows a choice, one for each graph symmetry, of stabilisers; we report only the optimal choice here, though all exceed the classical bound.} Other measurement results are reported in the Supplement. We find that $\ket{S_4}$ exceeds both $\Mtwo<2$ and $\Mthree<12$ classical bounds. $\ket{L_4}$ exceeds the classical bound for \Mtwo, but not for \Mthree, which is more strict. The higher postselection penalty of the controlled-$Z$, required to generate $\ket{L_4}$, results in a decreased fidelity, of which \Mthree is a simple rescaling.}

\begin{table*}
\begin{tabular*}{\textwidth}{l @{\extracolsep{\fill}} ccccc}
  State & Fidelity   & \Mtwo $(2,4)$ & \Mthree & Count rate & \josh{Counts}\vspace{1mm}\\
\hline
\hline\noalign{\vskip 1mm}    
 $\ket{S_4}$ & $0.78\pm 0.01$ & $3.17 \pm 0.07 $ & $12.45\pm 0.12  \quad (12, 16)$ & 5.7 mHz & $2640$\\ 
 $\ket{L_4}$ & $0.68\pm 0.02$ & $2.61\pm0.14 $ & $10.93\pm 0.29\quad (12, 16)$ & 1.1 mHz & $1085$\vspace{0.2cm}\\
 
 $\ket{S_4}_{1,2,4}$ & $0.77 \pm 0.01$ & $2.79 \pm 0.09  $ & $6.16 \pm 0.11 \quad (6, 8)$ & 3.3 mHz & $1142$\\
 $\ket{S_4}_{1,4}$ & $0.83 \pm 0.02$ & - & $3.32\pm 0.09 \quad (2, 4)$ & 4.0 mHz & $416$\\
 $\ket{S_4}_{3,4}$ & $0.83 \pm 0.02$ & - & $3.31\pm 0.09 \quad (2, 4)$ & 4.1 mHz & $369$\vspace{0.2cm}\\
 
 $\ket{\Phi^{+}}_{1,3}$ & $0.97 \pm 0.01$ & $2.79 \pm 0.01^\dagger$ & $3.90\pm 0.03 \quad (2, 4)$ & 1.8 kHz & $38003$\\
 $\ket{\Phi^{+}}_{2,4}$ & $0.97 \pm 0.01$ & $2.71 \pm 0.01^\dagger$ & $3.88\pm 0.03 \quad (2, 4)$ & 1.9 kHz & $41769$\\
 \end{tabular*}
\caption{Summary of measured parameters for on-chip graph states. State fidelities, Mermin test parameters, and photon statistics are listed. Classical and quantum bounds are listed in parentheses, where they apply. `$\dagger$' indicates a Bell-CHSH test.}
\label{tab-data}
\end{table*}

As quantum devices increase in complexity, the scaling of errors  is of critical importance. Error models differ substantially between platforms, even within optics. Here, we develop methods for quantifying \josh{low-level} performance parameters and apply them to our device. We seek to understand the effects of photon distinguishability, multiphoton contamination, and thermo-optic phase \josh{error}. Each effect is modelled independently. \josh{Since all effects contribute to the data, our estimates for each parameter are pessimistic.} We apply Bayesian parameter estimation to learn the likeliest model parameters based on the four-photon stabiliser data\cite{barber2012bayesian}. The indistinguishability ($\sigma$) , multiphoton emission ($p$), and random phase error ($\delta$), are estimated with no prior assumptions. The resulting probability distributions of the three parameters are reported in Fig.~\ref{fig-data}e--g, for both $\ket{S_4}$ and $\ket{L_4}$. Fitting each with a normal distribution, we compute parameter estimates and standard deviations: $\sigma_{S,L} = \{0.82 \pm 0.01, 0.82 \pm 0.01\}$, $p_{S,L} = \{0.036 \pm 0.009, 0.037 \pm 0.012\}$, and $\delta_{S,L} = \{0.185 \pm 0.007\ \mathrm{rad}, 0.182 \pm 0.009\ \mathrm{rad}\}$. Our other measurements (HOM interference, $g^{(2)}$, source brightness, and crosstalk---see Methods) are compatible with these estimates\josh{; the distributions for the two states, $\ket{S_4}$ and $\ket{L_4}$, also broadly agree}. This approach can reveal additional device performance information from \textit{existing} data---no new measurements are required.

To completely describe device performance a holistic error model---one that simultaneously captures all the effects---is needed. To formulate such a model requires knowledge of difficult-to-access quantities and significant computational power. \josh{A distinguishability model, for example, must have} the Schmidt spectrum of each source---inaccessible from simple HOM dips---and a common basis for them. Computationally, modelling variable, high photon-number states in high-dimensional spaces is a challenge. Moreover, the three effects we studied affected the observables in a similar way and depended on the state: a holistic model may not help to effectively distinguish these \josh{effects, but tailored or adaptive measurements may help.}

We have demonstrated a multiphoton, multiqubit capability using standard, commercially-available silicon photonic components. \josh{Six- or eight-photon devices can be built using the technique we show}: \josh{postselected} Bell-pair sources can be combined using the R-PEG gate to yield a plethora of graph states\cite{adcock2018hard} in the near future.

Though our postselection-reliant approach to sourcing photons and preparing entanglement is not scalable, scalable approaches (e.g. those using feedforward\cite{knill2001scheme, gimeno2015three, gimeno2017relative}) must overcome many of the same challenges. We can now bring the reconfigurability and control of integrated photonics to bear on the exploration of multiphoton space. The combination of multiple photons and high-dimensional techniques\cite{wang2018multidimensional} will soon make vast Hilbert spaces accessible. Postselection \josh{lets us test the components and techniques key to unlocking the huge} graph states needed for photonic quantum computation\cite{ gimeno2015three, rudolph2017optimistic}.

Graph states are, and will continue to be, a building block of large-scale quantum technology.
We have demonstrated a photonic generator of arbitrary graph states, in a miniature, high-performance technology.
For the first time, quantum information has been encoded in more than one pair of photons generated on a chip.
Future increases in photon number depend solely on engineering improved photon throughput.
This work lights the way towards a future of large-scale quantum photonic devices.

\section*{Acknowledgements}
This work was made possible with the support of Damien Bonneau, Chris Sparrow, Mercedes Gimeno-Sergovia, Sam Pallister, Will McCutcheon, Stefano Paesani, Eric Johnston, Laurent Kling, Graham D. Marshall, and John G. Rarity. This work was generously supported by EPSRC Programme Grant EP/L024020/1, the EPSRC Quantum Engineering Centre for Doctoral Training EP/L015730/1, and the ERC Starting Grant ERC-2014-STG 640079. JWS acknowledges the generous support of the Leverhulme Trust, through Leverhulme Early Career Fellowship ECF-2018-276. MGT acknowledges support from EPSRC Early Career Fellowship EP/K033085/1.

\section*{Author contributions}
JWS, RS, and JCA conceived the device. JCA and CV designed and carried out the experiment and experimental modelling. JWS and MGT supervised the project. All authors analysed the results and wrote the manuscript.

\section*{Methods}

\textbf{Experimental set-up.} Pump pulses at 1544.40~nm (1.1~ps pulse duration, 500~MHz repetition rate) from an erbium-doped fibre laser (Pritel) are filtered with square-shaped, $1.4$-nm-bandwidth filters and injected into the device. Signal and idler photons are collected at pump-detuned $\pm4.8$~nm, and filtered with square-shaped, 0.7-nm-bandwidth filters (Opneti DWDM) for spectral shaping and pump light rejection. They are detected off chip by four superconducting nanowire single photon detectors with $80\pm5\%$ efficiency (Photon Spot), operating around 0.85~K. Time-tags are generated (UQD-Logic) and converted to coincidences by bespoke software. The device is mounted using thermal epoxy and wire-bonded to an FR4 printed circuit board; temperature is stabilised using a closed-loop thermo-electric cooler. Optical coupling to fibre is via a fibre V-groove array (OZ Optics) and a 6-axis piezo-electric actuator (Thorlabs). Analogue voltage drivers (Qontrol Systems) are used to drive the on-chip phase shifters, with 16-bit and 300-$\upmu$V resolution. The device was fabricated by the A*STAR Institute of Microelectronics, Singapore. A 220-nm device layer performs waveguiding, atop a 2-$\um$ buried oxide (silicon-on-insulator) with an oxide top cladding. It has an area of $1.4 \times 3$~$\mathrm{mm}^2$ with 500-nm-wide waveguides. Kilohertz-bandwidth thermo-optic phase modulators are formed by TiN heaters, $180 \times 2\ \um^2$, positioned $2$~$\um$ above the waveguide layer. 

\vspace{1em}

\textbf{Phaseshifter calibration and crosstalk.} We calibrate the device's thermo-optic phaseshifters by illuminating their enclosing MZIs with a continuous-wave laser at the relevant wavelength, and applying a range of voltages to produce a fringe at the MZI output. We fit this fringe with a function $A \sin(f \cdot P(V) + \phi_0) + c$, where $P(V) = I(V)\cdot V$ is the Joule heating of the phaseshifter, to find $A$, $f$, $\phi_0$, and $c$. By measuring the current-voltage relationship of the phaseshifters and fitting them to $I (V) = \rho_1 V+ \rho_2V^2+ \rho_3 V^3$, we can `dial in' a phase $\phi_d$ by numerically solving the quartic equation  $\phi_d = f \cdot I(V)\cdot V + \phi_c$. Loss-matched, evanescently coupled waveguide taps with $ 2 $\% transmission are strategically placed around the device to allow independent calibration of each on-chip phaseshifter.

We measure the phase deviation within one on-chip demultiplexer per unit power dissipated in the other thermo-optic modulators. A thermal cross-talk coefficient of $ 0.003$~rad/mW results. The average power dissipated over all chip configurations used in the stabiliser measurements was $443$ mW and $472$ mW for the star and line states respectively. These distributions indicate an average deviation from the mean of $39$ mW and $22$ mW for the two states. Working backwards, we estimate the average thermo-optic phase error is $0.12$~rad and $0.065$~rad, respectively. \josh{Power histograms and cross-talk fringes are shown in the Supplementary Information.}

\vspace{1em}

\textbf{Loss.} The device insertion loss is 26.1~dB for the light path through source 1 to the $\ket 0$ output of qubit 1, after optimising the relevant phase settings. We estimate losses, based on measurements on test structures on the same die, as: 4~dB per vertical grating coupler, 0.65~dB per $2\times2$ multimode interferometer (MMI), 3~dB/cm of straight waveguide propagation, and 7.5~dB/cm of spiral waveguide propagation. All measurements are at 1544.4~nm. By including off-chip losses (3~dB),  input coupling (one grating, two MMIs), and one half of the source length, we estimate that signal photons experience a loss of 19.3~dB.

\vspace{1em}

\textbf{HOM-fringe visibilities.} In an ideal HOM \textit{fringe} the maximum is twice the background `distinguishable' level of an ideal HOM \textit{dip}. To calculate the equivalent \textit{dip} visibility $V_{\mathrm{HOM}}$ from the maximum and minimum values measured in a \textit{fringe}, we use $ V_{\mathrm{HOM}} = (N_\text{max}/2-N_\text{min}) / (N_\text{max}/2) = (N_\text{max}-2N_\text{min}) / N_\text{max} $. More details are in the Supplement.

\vspace{1em}

\textbf{Measuring state fidelities.} We wish to find the fidelity of our experimental state $\rho_{ex}$, with a graph state $ \rho$, with stabilisers $\{g_i\}$.  Since $\rho$ is a stabiliser state, $\rho = \frac{1}{2^n} \sum_i^{2^n} g_i $. Hence, $F= \text{tr}[\rho_{ex}\rho] = \frac{1}{2^n} \sum_i^{2^n} \text{tr}[g_i \rho_{ex}] = \frac{1}{2^n}\sum_i^{2^n} \langle g_i\rangle $ (see ref. \citenum{guhne2009entanglement}). This measurement method is used for all reported state fidelities.

Local Pauli expectation values are measured by projecting each of the $2^n$ eigenvectors onto \josh{each qubit's} single output waveguide and counting $n$-fold coincidences (in our experiment, $n=4$). Summing the results of each projective measurement (total counts $C_j$) by eigenvalue and normalising gives $\langle g_i \rangle  = \sum_j^{2^n} \lambda_j C_j /\sum_j^{2^n} C_j$. Here the eigenvalue of stabiliser projector $j$ is a product of its local components $\lambda_j = \prod_k^n\mu^{(k)}_j$, with $\mu_j^{(k)} \in \{-1,1\}$ being the eigenvalue of the local operator on qubit $k$. \jez{The Supplementary Information contains a complete list of each state's stabilisers}.

\vspace{1em}

\textbf{Mermin tests.} For both $\ket{S_4}$ and $\ket{L_4}$, we measure every two-setting Mermin test that \josh{can be} composed from its stabilisers. The tests for the star state are as follows (graph symmetries are indicated by an arrow): $\Mtos = g_4(1+g_2 g_3+g_2 g_1 +g_3 g_1), g_4 \rightarrow g_4 g_1$ and $\Mtosvar = g_4(1+g_i)(1+g_j), g_4\rightarrow g_4 g_k,$ where $g_i$ are the stabiliser generators and $i,j,k=\{1,2,3\}$. For the line state: $\Mtol = g_1(1+g_2)(1+ g_3),$ with $g_2\rightarrow g_2g_4$ and $\Mtolvar = g_1(1+g_3)(g_2+ g_4),$ with $g_2\rightarrow g_2g_4,$ and $g_i\rightarrow g_ig_{i+1},$ for $i\in\{1,2,3,4\}$. Local-realistic (``classical'') theories obey $|\langle \Mtwo \rangle|<2,$ while $|\langle \Mtwo \rangle|<4$ for quantum mechanics.

We also report a three-setting Mermin test: $\Mthree = \sum_{i} \langle g_i \rangle,$ where the sum is take over all the $2^n$ (16) stabilisers of the graph state. Local-realistic theories obey $|\langle \Mthree \rangle|<12,$ while $|\langle  \Mthree\rangle|<16$ for quantum mechanics.

\vspace{1em}

\textbf{Bayesian parameter estimation.} We use three \josh{independent models} to simulate the effects of partial distinguishability, multiphoton emission, and phase error (see the Supplementary Information for model details). These output a four-fold rate for each measurement setting, used to estimate a fidelity, for a range of $\sigma$, $p$, and $\delta$. The phase error model was based on $10^4$ normally distributed Monte Carlo samples for each chip configuration, with $\delta$ the phase offset standard deviation. Data from each model is compared to the experimentally obtained data, and Bayesian inference learns the likeliest value for each parameter.

Consider a system described by a known model $M(\sigma)$ with free parameter $\sigma$, a set of $N$ observables $\Pi = \{\pi_i\}_{i=1}^N$ and a data set $X = \{x_i\}_{i=1}^N$: the general aim of Bayesian parameter estimation is to find the parameter $\bar{\sigma}$  that best describes the data outputted by the system. Learning $\bar{\sigma}$ relies on the estimation of likelihoods, over a discretised space  $\{\sigma_k\}_{k=1}^K$ of $K$ possible $\sigma_k$s: $L(\sigma_k) = \prod_{i=1}^N P(x_i|\sigma_k, \pi_i),$ where $ P(x_i|\sigma_k, \pi_i) $ is the probability of observing $x_i$ given model parameter $\sigma_k$ and measured the observable $\pi_i$. This probability can be calculated from the frequency of the observed data $x_i$ over many samples of simulated data $\tilde{x_i}$. We can therefore derive the probability of $\sigma_k$ being the parameter that best describes the data by applying Bayes's rule:
\begin{align*}
P(\sigma_k|X,\Pi) &= \frac{P(X|\sigma_k, \Pi)  P(\sigma_k)}{\sum_{l=1}^K P(X|\sigma_l, \Pi) * P(\sigma_l)}\\
&= \frac{\prod_{i=1}^N P(x_i|\sigma_k, \pi_i) }{\sum_{l=1}^K \prod_{i=1}^N P(x_i|\sigma_l, \pi_i)} = \frac{L(\sigma_k)}{\sum_{l=1}^K L(\sigma_l)},
\end{align*}
thus retrieving a probability distribution for each parameter. We have assumed the measurements to be uncorrelated and the \textit{a priori} distribution of the parameters $P(\sigma_k)$ to be constant over the discretised range.

\end{document}